\documentclass[preprint2]{aastex631}

\accepted{December 11, 2022}

\submitjournal{AJ}

\begin{document}

\title{On averaging eccentric orbits: Implications for the long-term thermal evolution of comets}

\correspondingauthor{Anastasios Gkotsinas}
\email{anastasios.gkotsinas@univ-lyon1.fr}

\author[0000-0002-1611-0381]{Anastasios Gkotsinas}
\author[0000-0003-2354-0766]{Aur\'{e}lie Guilbert-Lepoutre}
\affiliation{Laboratoire de G\'{e}ologie de Lyon: Terre, Plan\`{etes}, Environnnement, CNRS, UCBL, ENSL, F-69622, Villeurbanne, France} 

\author[0000-0001-8974-0758]{Sean N. Raymond}
\affiliation{Laboratoire d'Astrophysique de Bordeaux, Univ. Bordeaux, CNRS, F-33615 Pessac, France}

\begin{abstract}

One of the common approximations in long-term evolution studies of small bodies is the use of circular orbits averaging the actual eccentric ones, facilitating the coupling of processes with very different timescales, such as the orbital changes and the thermal processing. Here we test a number of averaging schemes for elliptic orbits in the context of the long-term evolution of comets, aiming to identify the one that best reproduces the elliptic orbits' heating patterns and the surface and subsurface temperature distributions. We use a simplified thermal evolution model applied on simulated comets both on elliptic and on their equivalent averaged circular orbits, in a range of orbital parameter space relevant to the inner solar system. We find that time averaging schemes are more adequate than spatial averaging ones. Circular orbits created by means of a time average of the equilibrium temperature approximate efficiently the subsurface temperature distributions of elliptic orbits in a large area of the orbital parameter space, rendering them a powerful tool for averaging elliptic orbits.   

\end{abstract}

\keywords{}


\section{Introduction} \label{section:Intro}

Considering comets' long lifetimes and complex dynamical histories, studying the long-term evolution of their nuclei is an essential step in understanding their current state and activity \citep{Gkotsinas_2022}. One approach to assess long-term effects is to determine their thermal histories by coupling a thermal evolution model to orbital trajectories from dynamical simulations \citep{Raymond_2020, Gkotsinas_2022}. This coupling is challenging both from a physical and a numerical point of view \citep[][]{Gkotsinas_2022} and demands a certain number of assumptions, as the processes involved operate on very different timescales. Thermal evolution processes such as phase transitions, heat or gas diffusion, depending on the temperature conditions, usually take place in minutes or hours or up to a few months in the case of the heat diffusion \citep{Prialnik_2004}. The dynamical evolution of comets, on the other hand, is a billion-year process that requires N-body simulations with an output frequency on the order of hundreds to thousands of years \citep[e.g.][]{Nesvorny_2017, Sarid_2019}, far longer than the short timescales of thermal processes.

One of the assumptions frequently used in these type of simulations is to simplify the dynamical pathways of comets, by averaging their orbits \citep{Prialnik_2009, Guilbert-Lepoutre_2012, Snodgrass_2017, Gkotsinas_2022}. \citet{Prialnik_2009} proposed to replace elliptic orbits by equivalent circular ones; the radius of which was chosen to assure the same amount of total energy over the course of an orbital period. This is reinforced in other fields, as climate modelling of putative Earth-like planets on eccentric orbits has generally found that the controlling factor determining whether a planet may retain liquid water is the total energy received over a planet's orbit \citep{Williams_2002, Bolmont_2016}.

In this work we aim to compare and evaluate different methods for averaging elliptical orbits in the context of long-term simulations of comets' thermal evolution. In Section \ref{section:Methods} we present the tested averaging schemes and we give a brief presentation of the thermal evolution model. In Section \ref{section:Results} we present the produced internal temperature distributions from the different schemes and a comparative study between them and elliptic orbits. In Section \ref{section:Discussions} we discuss the adequacy and the limits of validity of every scheme and highlight the scheme that worked better in the current context.

\section{Methods} \label{section:Methods}

\subsection{Averaging eccentric orbits}

A variety of approaches have been proposed in order to average an elliptic orbit. The most common one is to estimate an average distance between the object and the focal point. This can be achieved in two ways: The first is to integrate the orbit equation over the true anomaly ($\theta$) throughout an orbital period to obtain a true-anomaly-averaged radius \citep[e.g.][]{Curtis_2014a}:

\begin{equation} \label{spatial_radius}
    \bar{r}_{\theta} = a \sqrt{1-e^2}
\end{equation}

The second way is to integrate the orbit equation over the average angular velocity of the object over the course of an orbital period. This approach results in a time-averaged radius, which is always smaller than the true-anomaly-averaged \citep[e.g.][p.145-186]{Curtis_2014a}: 

\begin{equation} \label{temporal_radius}
    \bar{r}_t = a \left( 1 + \frac{e^2}{2}   \right )
\end{equation}

Another widely used method is to integrate over a physical parameter, such as the flux received by an object over an orbital period. This is the most physically-plausible way to approach the problem, as it ensures that the total energy intercepted by an object over an orbit is not modified. Using Equations \ref{spatial_radius} and \ref{temporal_radius} we can calculate a true-anomaly-averaged and a time-averaged flux \citep{Mendez_2017} from which we can obtain a new set of radii, the second of which is commonly used on planetary habitability studies \citep[e.g.][]{Bolmont_2016}:

\begin{equation} \label{spatial_flux}
    \bar{r}_{\theta F} = \frac{a(1-e^2)}{\sqrt{2+e^2}}
\end{equation}

\begin{equation} \label{temporal_flux}
    \bar{r}_{tF} = a (1 - e^2)^{\frac{1}{4}}
\end{equation}

Recently, \citet{Mendez_2017} proposed a new effective thermal radius, calculated directly from the time average of the equilibrium temperature ($T_{eq}$), guaranteeing the same average equilibrium temperature over an orbital period:

\begin{eqnarray}
    r_T &=& a \left[ \frac{2 \sqrt{1+e}}{\pi} \textit{\textbf{E}} \sqrt{\frac{2e}{1+e}} \right]^{-2}\\
        &\approx& a (1 + \frac{1}{8}e^2 + \frac{21}{512}e^4 +\mathcal{O}(e^6) ) \label{eq_temperature}
\end{eqnarray} 

\noindent where \textit{\textbf{E}} is the complete elliptic integral of the second order.

All of the proposed expressions are simple functions of the semimajor axis ($a$) and the eccentricity ($e$). In practice, this means that for an elliptic orbit with specified orbital parameters (i.e. a ($a,e$) couple), the orbit-averaging technique produces a circular orbit around the focal point with an `equivalent' radius. The differences in the averaging expressions imply different distances from the Sun. In fact for any given ($a,e$) couple the calculated radii are ordered as: $\bar{r}_{\theta F} < \bar{r}_{\theta} < \bar{r}_{tF} < r_T < \bar{r}_t$. This means that true-anomaly-averaged flux expression ($\bar{r}_{\theta F}$) will always place an object closer to the Sun than the time-averaged radius ($\bar{r}_t$). As a consequence for the same ($a,e$) couple we produce different temperature profiles (panels (b) to (e) in Figure \ref{fig:T_profiles}), raising the question 'which one is better approximating the temperature distribution of the elliptic orbit (panel (a) in Figure \ref{fig:T_profiles})?'. To answer we test these orbit-averaging schemes in the context of comets' thermal evolution. In order to test the validity limits of each scheme in a range of orbital parameter space relevant for the inner solar system (i.e. $\sim$3-30 au) we use a total of 110 ($a,e$) couples. We sample the semimajor axis range logarithmically ($10^x$, with $x$ ranging between 0.5 and 1.5 with a step of 0.1) with more orbits close to the Sun, where the heating is stronger and the eccentricity linearly between 0 to 0.9 at increments of 0.1.

\subsection{Thermal evolution model}

We use a 1D version of the 3D thermal evolution model described in \citet{Guilbert_2011} to solve the heat diffusion equation in a spherical airless object:

\begin{equation}
    \rho_{bulk}c~ \frac{\partial T}{\partial t} + div(-\kappa~ \overrightarrow{grad} T) = \mathcal{S} \label{eq:heat_diffusion}
\end{equation}

\noindent where $\rho_{bulk}$ is the object's bulk density (kg m$^{-3}$), $c$ the material's heat capacity (J kg$^{-1}$ K$^{-1}$), $T$ the temperature (K), $\kappa$ the material's effective thermal conductivity (W K$^{-1}$ m$^{-1}$), and $\mathcal{S}$ the heat sources and sinks.

The surface boundary condition for Equation \ref{eq:heat_diffusion} is:
\begin{equation}
    (1-\mathcal{A}) \frac{L_\odot}{4\pi d_H^2} = \varepsilon \sigma T^4 + \kappa \frac{\partial T}{\partial z} \label{boundary_eq}
\end{equation}
\noindent where the received solar energy is given as a function of the Bond's albedo ($\mathcal{A}$), the solar constant ($L_\odot$) in W m$^{-2}$ and the heliocentric distance ($d_H$) in au; the nucleus' thermal emission as a function of the emissivity ($\varepsilon$), the Stefan-Boltzmann constant ($\sigma$) and the temperature (T) in K; and the heat flux toward the interior given as a function of the surface's thermal conductivity ($\kappa$) in W~K$^{-1}$~m$^{-1}$.
We assume that the incident solar energy is uniformly distributed over the surface of the sphere, providing a spherical average of the energy received by the nucleus.

As the goal of this work is to compare elliptic to circular orbits, we chose a simplified setup for our model. Each comet is composed of dust without any ice, such that no phase transitions take place, removing any energy sources or sinks from Equation \ref{eq:heat_diffusion} ($\mathcal{S}$=0). Without them the most important parameter in the model is the effective thermal conductivity ($\kappa$), as it controls the heat diffusion toward the interior. In the current configuration it is set at 5$\times$10$^{-3}$ W m$^{-1}$ K$^{-1}$, in good agreement with laboratory measurements for porous dust aggregates \citep{Krause_2011}. The rest of the model's parameters are widely used averages in the published literature \citep{Huebner_2006}. 

It is the value of the heliocentric distance ($d_H$) that changes between the different case studies, controlling the amount of energy received at the surface of our objects. In an elliptic orbit it changes following the constant increment of the eccentric anomaly as the object moves between the apsides. In the test cases it is constant, set on the distance calculated by Equations \ref{spatial_radius}-\ref{eq_temperature}.

We run a total of 660 simulations: 110 reference simulations for all the ($a,e$) couples with objects on elliptic orbits, serving as basis for the comparisons with the 550 simulations for objects on equivalent circular orbits created from Equations \ref{spatial_radius} to \ref{eq_temperature} for the same ($a,e$) couples. The simulations run for $\sim$1 Myr, an arbitrary period selected to allow heat diffusion inside our objects. This allows to study the temperature differences between elliptic and circular orbits not only at the surface but in the interior as well, and look for any accumulative or propagation effects that might be introduced during long-term simulations.

\section{Results} \label{section:Results}

\begin{figure}
    \centering
    \includegraphics[width=\columnwidth]{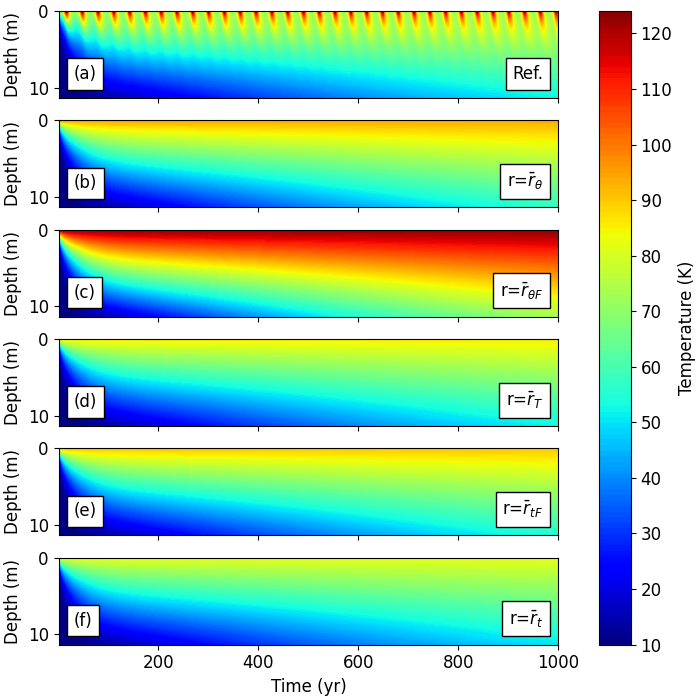}
    \caption{Subsurface temperature distributions for a layer of 10 m over a period of 1000 years for: (a) an elliptical orbit with $a$=10 au and $e$=0.5, (b) true-anomaly-averaged radius ($\bar{r}_{\theta}$), (c) true-anomaly-averaged flux ($\bar{r}_{\theta F}$), (d) effective thermal radius ($r_T$), (e) time-averaged flux ($\bar{r}_{tF}$) and (f) time-averaged radius ($\bar{r}_{t}$) equivalent circular orbits.}
    \label{fig:T_profiles}
\end{figure}

Figure \ref{fig:T_profiles} presents the internal temperature distribution produced by the different orbit-averaging schemes for a comet on an orbit with $a$=10 au and $e$=0.5. For clarity only the first 1000 yr of the 1 Myr simulation are presented, but this is sufficient to notice the averaging effects. Clearly, none of the averaged orbits reproduce the heating cycle of the elliptic orbit with the subsequent passages from the perihelion to the aphelion, i.e. the seasons. Instead, as the distance from the Sun is constant, the heat diffusion is steady and uniform throughout an orbital period.

\begin{figure*}
    \centering
    \includegraphics[width=\textwidth]{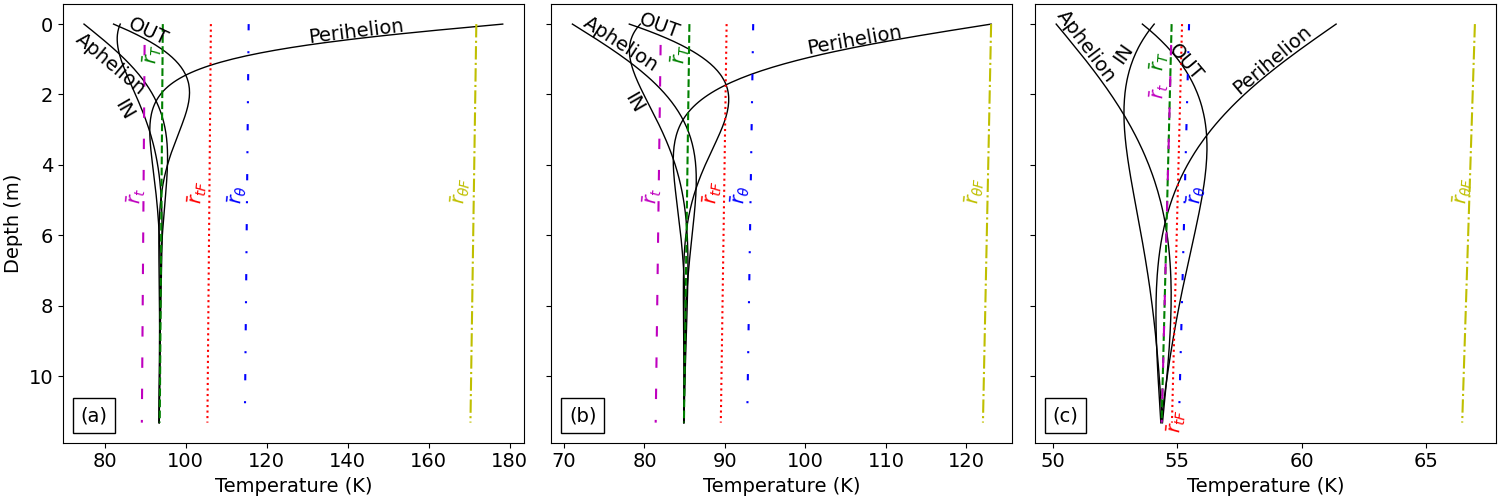}
    \caption{Temperature profiles for a subsurface layer of 10 m, $\sim$1 Myr after the start of the simulations, for three $a,e$ couples: (a) $a$=7.94 au and $e$=0.7, (b) $a$=10.0 au and $e$=0.5, (c) $a$=25.11 and $e$=0.2. The solid black lines give the temperature profiles of elliptic orbits at perihelion, aphelion and halfway through -time-wise- both inwards and outwards. The temperatures profiles for the equivalent orbits are: true-anomaly-averaged radius ($\bar{r}_{\theta}$) (blue loosely dashed-dotted line), true-anomaly-averaged flux ($\bar{r}_{\theta F}$) (yellow dashed-dotted line), time-averaged radius ($\bar{r}_{t}$) (green dashed line), time-averaged flux ($\bar{r}_{tF}$) (red dotted line) and effective thermal radius ($r_T$) (purple loosely dashed line).}
    \label{fig:T_profiles_lines_panels}
\end{figure*}

Figure \ref{fig:T_profiles_lines_panels} presents temperature profiles in the interior of our simulated comets using different orbit-averaging methods, taken near the end of our simulations ($\sim$1 Myr). The fact that the profiles are near-vertical stems from the assumption of a fixed, circular orbit with constant illumination. A profile from the reference elliptic orbit with the same orbital elements (i.e. $a$ and $e$) is shown at perihelion and aphelion, and at two points halfway in between at times coinciding with a quarter of its orbital period, both on its way inwards (from aphelion to perihelion) and outwards (from perihelion to aphelion). The three panels examine different orbits: (a) a highly eccentric and relatively short orbit with $a$=7.94 au and $e$=0.7, (b) a longer, less eccentric orbit with $a$=10 au and $e$=0.5 and (c) a long orbit ($a$=25.11 au) with low eccentricity of 0.2.

The temperature profiles from Figure \ref{fig:T_profiles_lines_panels} help us to confirm some of the previous observations:
\begin{itemize}
    \item The true-anomaly-averaged flux ($\bar{r}_{\theta F}$) better approximates the effect of the perihelion passage (even for the less eccentric orbit, where the temperature difference at the surface is $\sim$6 K), but fails completely on the other positions both on the surface and the subsurface: Divergences range from $\sim$6 K in the least eccentric to $\sim$80 K in the most eccentric orbit examined.
    \item The true-anomaly-averaged radius ($\bar{r}_{\theta}$) works slightly better, with calculated surface temperatures between those of the two apsides and subsurface profiles closer to the reference ones, and with temperature differences ranging from $\sim$20 K in the most eccentric, to almost complete convergence in the least eccentric orbits.
    \item Overall, the time-averaged expressions (Equations \ref{temporal_radius}, \ref{temporal_flux} and  \ref{eq_temperature}) work better in all cases.  Despite their failure to reproduce the high surface temperatures encountered at perihelion, especially for the most eccentric orbit, these schemes better reproduce the cooling effect of the aphelion passage, leading to internal temperature distributions that converge to those of the reference orbit in all the examples of Figure \ref{fig:T_profiles_lines_panels}.
    \item The effective thermal radius ($r_T$) stands out as the scheme producing an internal temperature distribution converging almost perfectly to the reference distribution, close to the surface: below 6 m in the first two cases and 10 m in the last.
\end{itemize}

Given that the true-anomaly-averaged (or spatial) expressions failed in matching the reference simulations, we focus hereafter on the time-averaged formulas. We expand our study by examining the temperature differences from the reference elliptic orbits at the surface, 1 and 10 m below the surface at perihelion, at aphelion and halfway in between time-wise. 

\begin{figure*}
    \centering
    \includegraphics[width=\textwidth]{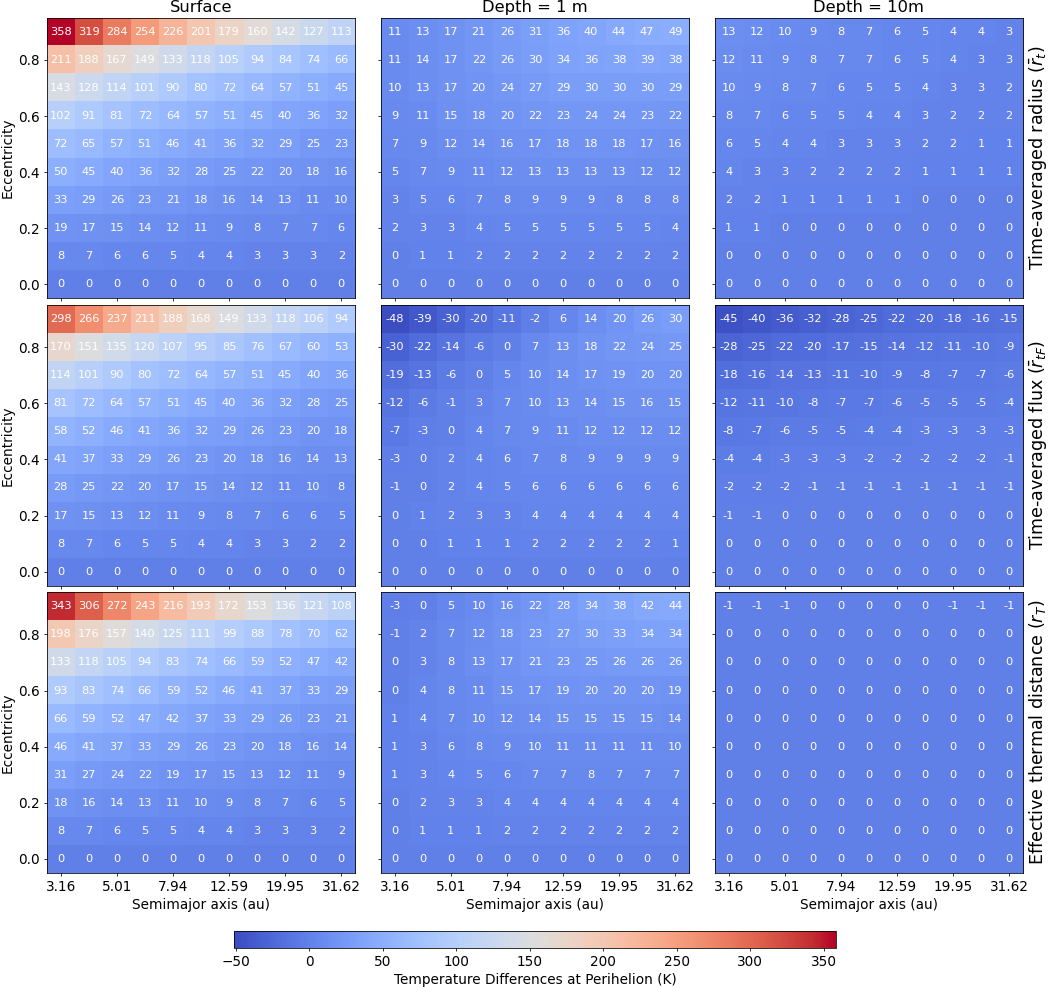}
    \caption{Temperature differences between the elliptic orbits at perihelion and the time-averaged radius ($\bar{r}_{t}$) (top row), the time-averaged flux ($\bar{r}_{tF}$) (middle row) and the effective thermal radius ($r_T$) equivalent orbits (bottom row) for three depths: surface (left panels), 1 m (middle panels) and 10 m (right panels) for all the $a,e$ couples.}
    \label{fig:taF_taR_heatmap_at_q}
\end{figure*}

In Figure \ref{fig:taF_taR_heatmap_at_q} we present the temperature differences ($\Delta T$) between the three time-averaged formulas ($\bar{r}_t$, $\bar{r}_{tF}$, $r_T$) and the elliptic orbits at perihelion. All schemes underestimate the surface temperatures during the perihelion passage. The time-averaged radius ($\bar{r}_t$) deviates the most from the reference orbits (maximum $\Delta T$ of $\sim$358 K or in terms of relative difference by 73\%), especially for very eccentric ($e>$0.5) and short orbits ($a<$10 au). It is followed by the effective thermal radius ($r_T$) with maximum divergence of $\sim$343 K (relative difference of 70\%). The time-averaged flux ($\bar{r}_{tF}$) presents a maximum $\Delta T$ of $\sim$298 K (or relative difference of 61\%). These relative differences at the surface rise with the eccentricity, becoming important ($\sim$20\%) above $e$=0.3, and really significant for high eccentricities (50\% for $e$=0.7 and $\sim$70\% for $e$=0.9) for the time-averaged and the effective thermal radius, highlighting their failure to represent the perihelion passage. The time-averaged flux ($\bar{r}_{tF}$) has slightly smaller relative deviations especially above $e>$0.4 where it is constantly lower than the other time-averaged schemes by $\sim$4-9\%. 

All three schemes are more robust in the objects' interiors (middle- and right-hand panels of Figure \ref{fig:taF_taR_heatmap_at_q}). At 1 m below the surface the time-averaged radius ($\bar{r}_t$) differences converge quickly for all orbits (maximum $\Delta T$ of $\sim$26 K) except for distant ($a>$10 au) and highly eccentric ($e>$0.5) orbits for which the maximum $\Delta T$ is $\sim$49 K. The same stands for the effective thermal radius ($r_T$) only with better convergence for short ($a<$10 au) and low-eccentric orbits ($e<$0.5). For distant ($a>$10 au) and highly eccentric ($e>$0.5) orbits the problem remains but is slightly less pronounced (with maximum $\Delta T$ of $\sim$44 K). Long period orbits with prolonged excursions into hot areas around perihelion allow the heatwave to advance deeper in the interior and extend the temperature differences well below the surface. In these cases underestimating the perihelion temperature remains problematic, unlike the cases of short orbits ($a<$10 au) where the heating, although more intense, takes place in a shorter period of time that is not sufficient for its diffusion in the interior (panel (a) versus panel (b) in Figure \ref{fig:T_profiles_lines_panels} for example). These differences almost completely disappear 10 m below the surface in the case of the effective thermal radius ($r_T$) ($\Delta T_{max}$=-1 K for $e=$0.9, lower right panel in Figure \ref{fig:taF_taR_heatmap_at_q}) and the time-averaged radius ($\bar{r}_{t}$) (with the exception of very short ($a<$10 au ) and highly eccentric orbits ($e>$0.7)). On the contrary, the time-averaged flux ($\bar{r}_{tF}$) differences at the interior (middle panels in Figure \ref{fig:taF_taR_heatmap_at_q}) fail to achieve convergence in high eccentricities ($e\geq$0.5), whether it is a short or a long orbit. In addition these deviations persist at larger depths (middle right panel of Figure \ref{fig:taF_taR_heatmap_at_q}), as there is still no convergence for highly eccentric orbits ($e\geq$0.5) and the differences on low eccentricity orbits are higher than those of the effective thermal radius and the time-averaged radius. 

\begin{figure*}
    \centering
    \includegraphics[width=\textwidth]{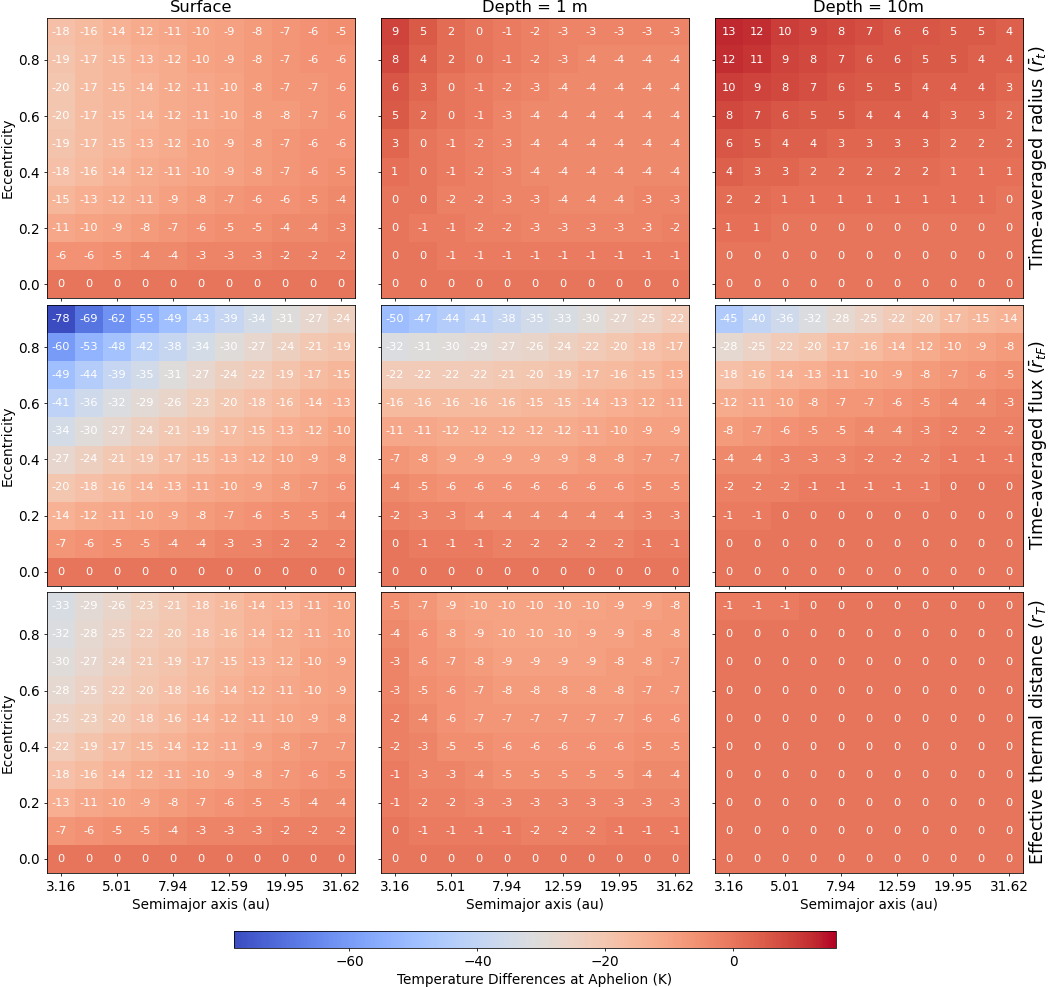}
    \caption{Temperature differences between the elliptic orbits at aphelion and the time-averaged radius ($\bar{r}_{t}$) (top row), the time-averaged flux ($\bar{r}_{tF}$) (middle row) and the effective thermal radius ($r_T$) equivalent orbits (bottom row) for three depths: surface (left panels), 1 m (middle panels) and 10 m (right panels) for all the $a,e$ couples.}
    \label{fig:taF_taR_heatmap_at_Q}
\end{figure*}

At the aphelion the temperatures are overestimated by the averaging schemes (Figure \ref{fig:taF_taR_heatmap_at_Q}). The time-averaged radius ($\bar{r}_t$) works better than the other methods in the surface and the close subsurface area ($\Delta T_{max}$=20 K versus $\Delta T_{max}$=-78 K for the $\bar{r}_{tF}$ and $\Delta T_{max}$=33 K for the $r_T$ at the surface and $\Delta T_{max}$=9 K versus $\Delta T_{max}$=50 K  and $\Delta T_{max}$=10 K respectively, 1 m below the surface). However, at 10 m below the surface the convergence for the effective thermal radius is almost complete ($\Delta T_{max}$=-1 K in only three short and eccentric orbits) unlike for the time-averaged radius and the time-averaged flux where there is still no convergence. For completeness, similar representations for the temperature differences halway through inwards and outwards presented in Figure \ref{fig:T_profiles_lines_panels} are given in Figures \ref{fig:Midway_IN} and \ref{fig:Midway_OUT}. As expected the temperature differences are less important comparing to those of the two apsides, but we can still observe the efficiency of the time-averaged and especially of the effective thermal radius over the time-averaged flux method. 

\section{Discussion} \label{section:Discussions}

Overall the time-averaged schemes work better than the true-anomaly or spatial-averaged ones. This is because the true-anomaly formulas are restricted to the calculation of an average distance from the focal point. Although this seems to be a sufficient assumption, it ignores a crucial information: the different time spent by an object at different distances from the focal point. In fact an object in an elliptic orbit will move much faster close to perihelion than close to aphelion, implying -in our case- more time in colder regions. This information is integrated in temporal expressions rendering them more appropriate in the approximation of elliptic orbits. With that in mind, the surface temperatures obtained from the time-averaged schemes, closer to the aphelion temperatures of an elliptic orbit, are more appropriate than an average of the perihelion-aphelion temperatures calculated by the spatial-averaged schemes.

When it comes to the time-averaged orbits, we demonstrated (Figures \ref{fig:taF_taR_heatmap_at_q}, \ref{fig:taF_taR_heatmap_at_Q}) that the time-averaged ($\bar{r}_t$) and the effective thermal radius ($r_T$) better approximate the temperature distributions of elliptic orbits with the exception of the surface temperatures at perihelion. These two distances are increasing functions of the eccentricity whereas the time-averaged flux ($\bar{r}_{tF}$) is a decreasing one \citep{Mendez_2017}. This implies that for a given ($a,e$) couple, the time-averaged flux will place an object closer to the perihelion, leading to a systematic overestimation of the surface temperature. On the other hand, the time-averaged and the effective thermal radius, (very close by definition, see Equations \ref{temporal_radius} and \ref{eq_temperature}) place the object closer to the aphelion accounting better for the lower temperatures reigning during the biggest part of an orbit (especially for highly eccentric ones), managing better to represent the internal temperature distribution. Interestingly, there is no clear distinction on the efficiency between these two formulas, as the time-average radius, as expected by its definition, works better at aphelion and halfway through both inwards and outwards at the surface and 1 m below, but fails to convergence as quickly as the effective thermal radius which manages to converge in all cases at maximum 10 m below the surface. 

We tested the equivalent semimajor axis proposed by \citet{Prialnik_2009}, deriving also from a time-averaging integral ($a_c = a(1-e^2)$). As this average distance provides the same energy per orbit as the real eccentric one, it is reliable only when orbital periods are very close. Otherwise significant deviations are observed (for very eccentric orbits for instance), with a systematic overestimation of the internal temperatures. Its validity thus remains limited to low eccentricity orbits ($e<$0.3) or limited timescales, as done in \citet{Guilbert-Lepoutre_2012, Snodgrass_2017, Gkotsinas_2022}.

\begin{figure}
    \centering
    \includegraphics[width=\columnwidth]{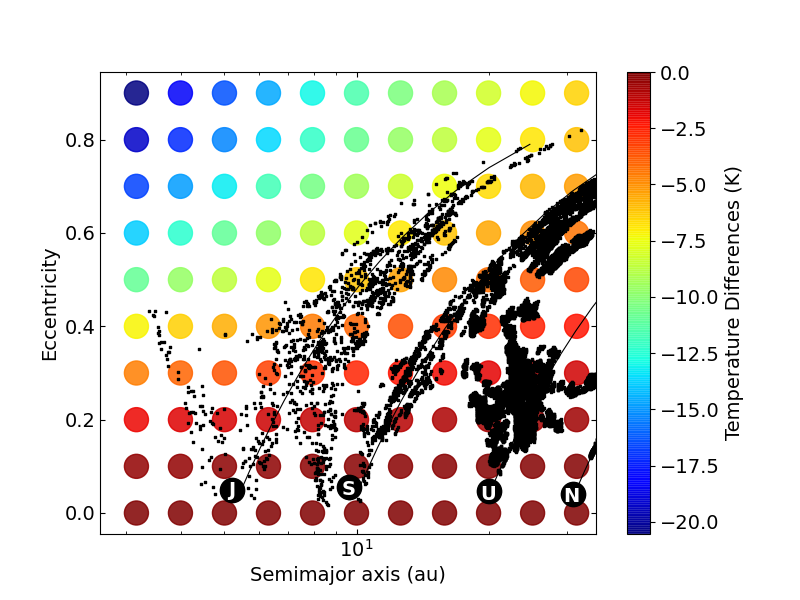}
    \caption{A typical Jupiter-family Comet trajectory taken from \citet{Gkotsinas_2022} plotted over the temperature differences between elliptic orbits and an average of the effective thermal radius ($r_T$) inwards and outwards for all the $a,e$ couples. Each circle represents a point in our orbital parameter space sampling. The color code gives the scale of the temperature difference.}
    \label{fig:Overplot}
\end{figure}

When the actual orbital trajectory of a typical comet is accounted for, it is clear that orbit-averaging remains a viable technique for the long-term cometary thermal evolution. Figure \ref{fig:Overplot} shows the orbital changes of a simulated Jupiter-family Comet during its trajectory toward the inner solar system (taken from \citet{Gkotsinas_2022} with dynamical trajectories from \citet{Nesvorny_2017}), over an average of the inwards and outwards surface temperature differences for the effective thermal radius ($r_T$). With the exception of short and highly eccentric orbits ($a<$10 au and $e>$0.5), which are very rare, the averaging is very efficient in all the other areas of the orbital parameter space. Even in the area of long and highly eccentric orbits ($a>$10 au and $e>$0.5), the temperature divergences do not overcome the $\sim$10 K. This suggests that using the effective thermal radius to average elliptic orbits is a very effective tool with a large area of validity in the orbital parameter space and not important discrepancies outside of this area (as even for the short and highly eccentric orbits the divergence is not bigger than 20 K). 

We therefore recommend the use of the effective thermal radius scheme: $r_T \approx a (1 + \frac{1}{8}e^2 + \frac{21}{512}e^4 +\mathcal{O}(e^6) )$, in long-term thermal evolution studies, as it can efficiently approximate the internal temperature distribution of an airless dusty body with low thermal inertia, such as a comet or an asteroid, evolving in eccentric orbits.

\begin{acknowledgments}
This study is part of a project that has received funding from the European Research Council (ERC) under the European Union’s Horizon 2020 research and innovation program (Grant agreement No. 802699). We gratefully acknowledge support from the PSMN (Pôle Scientifique de Modélisation Numérique) of the ENS de Lyon for computing resources. The authors would like to thank Benoit Carry for his useful comments on previous work, which provided the background for the current paper. 
\end{acknowledgments}

\vspace{5mm}
\facilities{PSMN, ENS de Lyon}

\bibliography{AveragedOrbits}{}
\bibliographystyle{aasjournal}

\appendix

\section{Supplementary figures}

\begin{figure*}[!h]
    \centering
    \includegraphics[width=\textwidth]{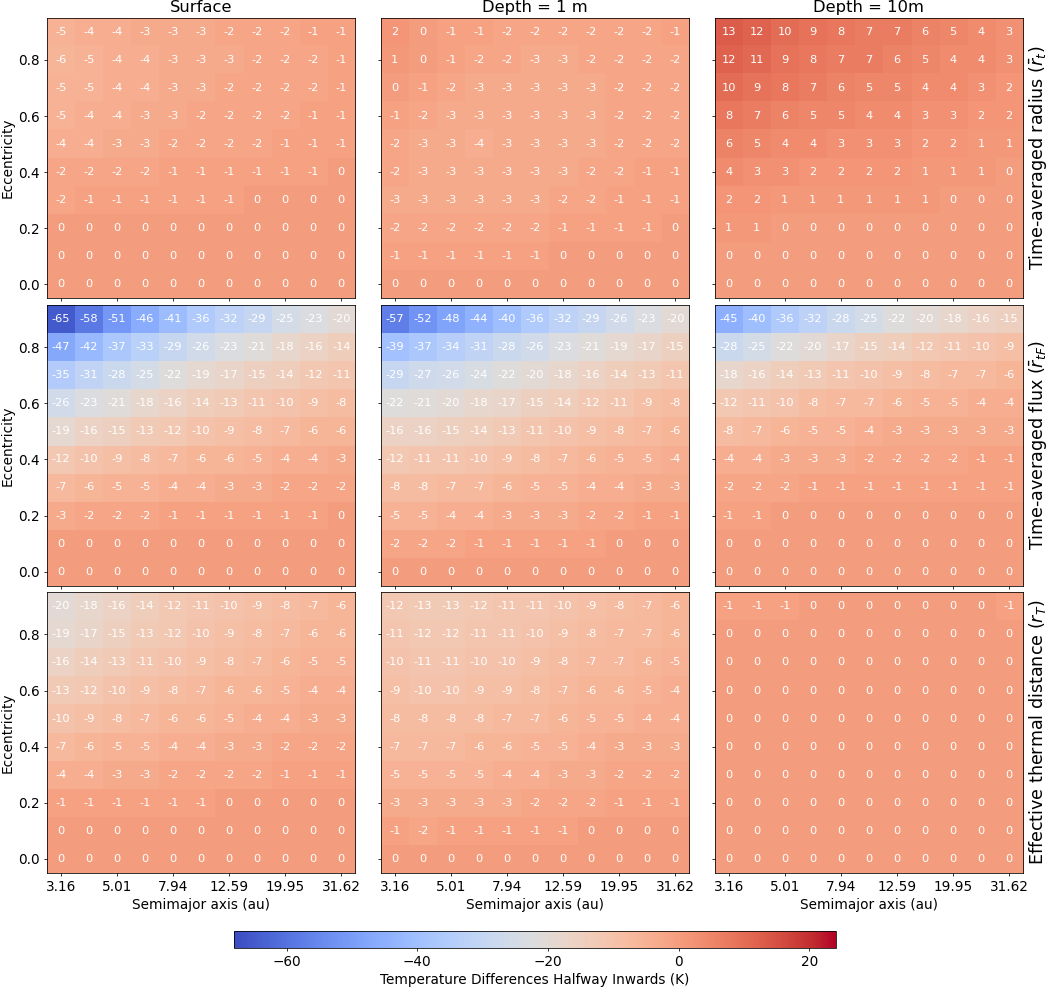}
    \caption{Temperature differences between the elliptic orbits halfway through inwards (time-wise) and the time-averaged radius ($\bar{r}_{t}$) (top row), the time-averaged flux ($\bar{r}_{tF}$) (middle row) and the effective thermal radius ($r_T$) equivalent orbits (bottom row) for three depths: surface (left panels), 1 m (middle panels) and 10 m (right panels) for all the $a,e$ couples.}
    \label{fig:Midway_IN}
\end{figure*}

\begin{figure*}[!h]
    \centering
    \includegraphics[width=\textwidth]{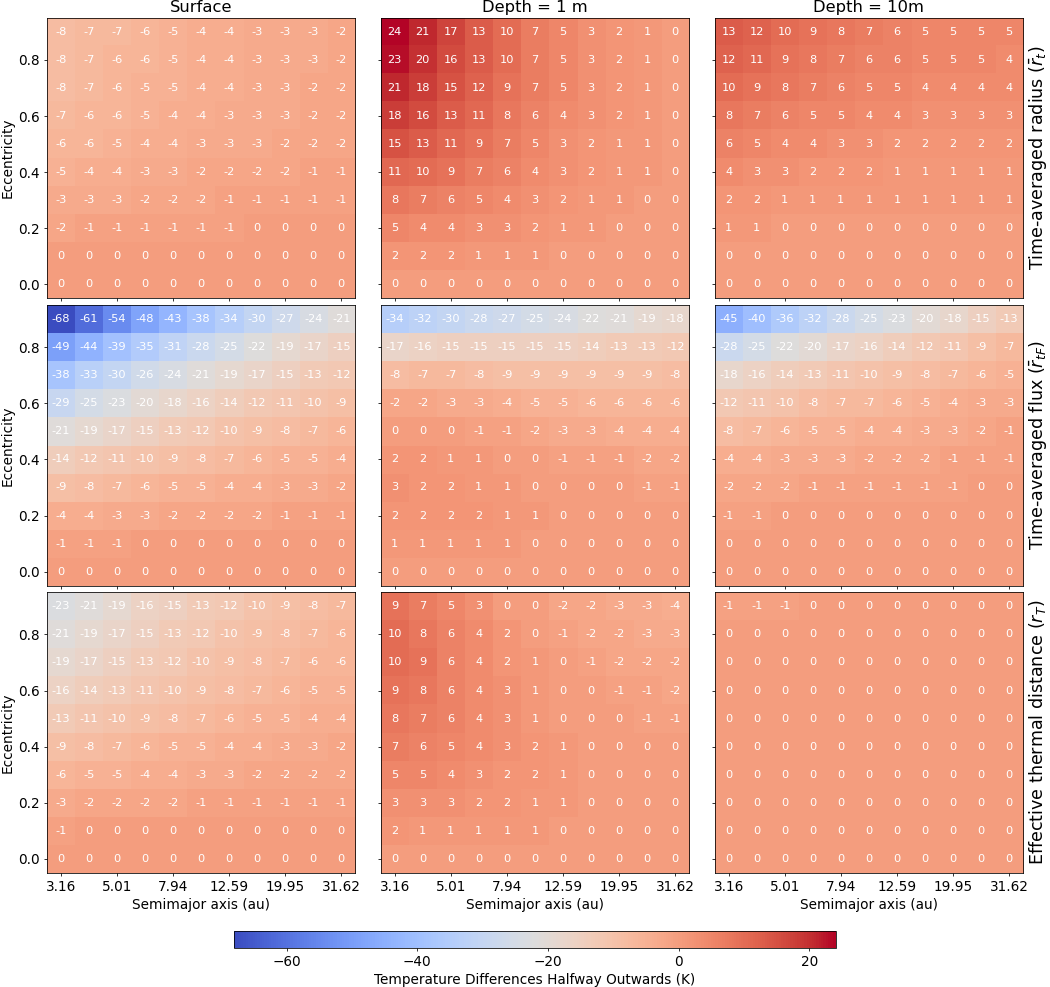}
    \caption{Temperature differences between the elliptic orbits halfway through outwards (time-wise) and the time-averaged radius ($\bar{r}_{t}$) (top row), the time-averaged flux ($\bar{r}_{tF}$) (middle row) and the effective thermal radius ($r_T$) equivalent orbits  (bottom row) for three depths: surface (left panels), 1 m (middle panels) and 10 m (right panels) for all the $a,e$ couples.}
    \label{fig:Midway_OUT}
\end{figure*}

\end{document}